# Maslov index for power-law potentials

**A. M. Ishkhanyan[a, b] and V. P. Krainov[c]**

[a] *Institute for Physical Research, National Academy of Sciences of Armenia, Ashtarak, 0203 Armenia*
[b] *Institute of Physics and Technology, National Research Tomsk Polytechnic University, Tomsk, 634050 Russia*
[c] *Moscow Institute of Physics and Technology (State University), Dolgoprudnyi, Moscow region, 141700 Russia*

The Maslov index in the semiclassical Bohr–Sommerfeld quantization rule is calculated for one-dimensional power-law potentials $V(x) = -V_0/x^s$, $x > 0$, $0 < s < 2$. The result for the potential $V(x) = -V_0/x^{1/2}$ is compared with the recently reported exact solution. The case of a spherically symmetric power-law potential is also considered.

## 1. INTRODUCTION

In the simplest one-dimensional case, the Maslov index $\gamma$ [1, 2] represents a correction to the quantum number $n$ that enters the Bohr–Sommerfeld quantization rule (hereinafter, $\hbar = m = 1$):

$$\int_a^b \sqrt{2(E_n - V(x))}\,dx = \pi(n + \gamma). \quad (1)$$

Such a correction is physically meaningful: although $n \gg 1$, the accuracy of the semiclassical approximation is higher, i.e., $1/n^2 \ll 1$. It is well known that, for potentials without singularities in the classically accessible region $a < x < b$, this correction is $\gamma = 1/2$ [3]. The Maslov index differs from 1/2 when the potential is singular at the turning points or elsewhere. In the general case, the Maslov index evidently falls within the range $-1 < \gamma < 1$. In this study, we calculate this correction for one-dimensional potentials of the form

$$V(x) = \begin{cases} -V_0/x^s, & x > 0, \\ \infty, & x < 0 \end{cases} \quad (2)$$

and for similar spherically symmetric potentials. Here, $0 < s < 2$ (for $s > 2$, the moving particle falls to the origin [3]: all discrete levels "sink" to minus infinity). Without limiting the generality, it may be assumed that $V_0 = 1$. The potential described by Eq. (2) is singular at the origin. The effect of this singularity is known only for the case $s = 1$ corresponding to the Coulomb potential, for which $\gamma = 0$ [4].

## 2. METHOD OF CALCULATION

We consider the Schrödinger equation

$$-\frac{1}{2}\frac{d^2\psi}{dx^2} - \frac{1}{x^s}\psi = E_n\psi; \quad (3)$$
$$x > 0, \quad n \gg 1, \quad |E_n| \ll 1.$$

In the region $x \ll x_0$, where $x_0 = |E_n|^{-1/s} \gg 1$ is the right turning point, the right-hand side of Eq. (3) can be disregarded, and the equation assumes the form

$$\frac{d^2\psi}{dx^2} + \frac{2}{x^s}\psi = 0. \quad (4)$$

The condition of applicability of the semiclassical approximation

$$\frac{d\lambda}{dx} \sim x^{s/2-1} \ll 1 \quad (5)$$

is violated for $x \ll 1$; so, the exact solution of Eq. (4) has to be found. The semiclassical approximation for Eq. (4) is valid only in the region of $1 \ll x \ll x_0$, where the corresponding solution can be written as

$$\psi(x) \sim x^{s/4} \cos\left(\int_0^x p(x')dx' - \delta\right)$$
$$= x^{s/4} \cos\left(\frac{2^{3/2}}{2-s} x^{1-s/2} - \delta\right). \quad (6)$$

This suggests seeking the exact solution of Eq. (4) by making the substitution of the independent variable $z = \frac{2^{3/2}}{2-s} x^{1-s/2}$. Then, we obtain

$$\frac{d^2\psi}{dz^2} - \frac{\alpha}{z}\frac{d\psi}{dz} + \psi = 0; \quad \alpha = \frac{s}{2-s}. \quad (7)$$

Making now the substitution $\psi(z) = z^{(1+\alpha)/2}\varphi(z)$, we reduce Eq. (7) to the Bessel equation

$$\frac{d^2\varphi}{dz^2} + \frac{1}{z}\frac{d\varphi}{dz} + \left(1 - \frac{\nu^2}{z^2}\right)\varphi = 0, \quad \nu = \frac{\alpha+1}{2}. \quad (8)$$

The solution of Eq. (8) satisfying the condition $\varphi(0) = 0$ is $\varphi(z) \sim J_\nu(z)$. For $z \gg 1$, we arrive at the asymptotic expression

$$\varphi(z) \sim \frac{1}{\sqrt{z}} \cos\left(z - \frac{\nu\pi}{2} - \frac{\pi}{4}\right), \quad (9)$$

so that the wavefunction asymptotically assumes the following form for $1 \ll x \ll x_0$:

$$\psi(x) \sim x^{s/4} \cos\left(\frac{2^{3/2}}{2-s} x^{1-s/2} - \frac{s\pi}{4(2-s)} - \frac{\pi}{2}\right). \quad (10)$$

Comparing Eqs. (6) and (10), we find

$$\delta = \frac{s\pi}{4(2-s)} + \frac{\pi}{2}. \quad (11)$$

Now, let us consider the semiclassical wavefunction in the region $1 \ll x < x_0$, starting from the right turning point:

$$\psi(x) \sim \frac{1}{\sqrt{p(x)}} \cos\left(\int_x^{x_0} p(x')dx' - \frac{\pi}{4}\right),$$

$$p(x) = \sqrt{2\left(\frac{1}{x^s} - \frac{1}{x_0^s}\right)}. \quad (12)$$

Writing $\int_x^{x_0}\ldots = \int_0^{x_0}\ldots - \int_0^x\ldots$, we obtain the following expressions for the integrals in Eq. (12) for the region $1 \ll x \ll x_0$:

$$\int_0^{x_0} \sqrt{2\left(\frac{1}{x^s} - \frac{1}{x_0^s}\right)} dx = \frac{\sqrt{2\pi}}{2s}\frac{\Gamma\left(\frac{1}{s} - \frac{1}{2}\right)}{\Gamma\left(\frac{1}{s}+1\right)}|E_n|^{1/2-1/s}, \quad (13)$$

$$\int_0^x \sqrt{2\left(\frac{1}{x'^s} - \frac{1}{x_0^s}\right)} dx' \approx \frac{2^{2/3}}{2-s} x^{1-s/2}. \quad (14)$$

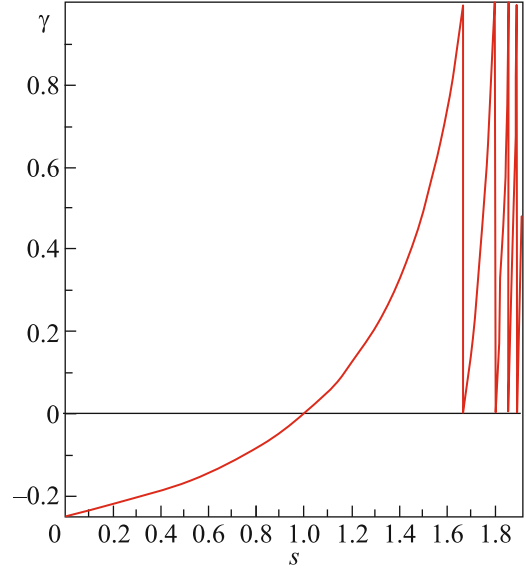

Maslov index for power-law potentials of the form $-1/x^s$.

Then, the wavefunction given by Eq. (12) assumes the following form in the region $1 \ll c \ll x_0$:

$$\psi(x) \sim x^{s/4}$$

$$\times \cos\left(\sqrt{\frac{\pi}{2}}\frac{\Gamma\left(\frac{1}{s} - \frac{1}{2}\right)}{\Gamma\left(\frac{1}{s}\right)}|E_n|^{1/2-1/s} - \frac{2^{3/2}}{2-s} x^{1-s/2} - \frac{\pi}{4}\right). \quad (15)$$

The Bohr–Sommerfeld quantization rule is obtained from the condition that wavefunctions (10) and (15) coincide with each other up to a phase factor; i.e., the sum of the cosine arguments has to be a multiple of $\pi$:

$$E_n = -\left\{\frac{\sqrt{2\pi}\cdot\Gamma\left(\frac{1}{s}\right)}{\Gamma\left(\frac{1}{s}-\frac{1}{2}\right)}\left(n + \left[\frac{s-1}{2(2-s)}\right]\right)\right\}^{\frac{2s}{s-2}}. \quad (16)$$

Thus, we find the Maslov index

$$\gamma = \left[\frac{s-1}{2(2-s)}\right]. \quad (17)$$

The square brackets in Eqs. (16) and (17) designate the fractional part of a number. In particular, for the case of $s = 1$ (corresponding to the Coulomb potential), we have $\gamma = 0$ and $E_n = -1/2n^2$. The plot of Eq. (17) is shown in the figure.

The ground-state energy for the case $s = 1$ is $-1/2$, as should be for the Coulomb potential. For $s = 0$, the ground-state energy is $-1$, in agreement with Eq. (22) below. For $s = 1/2$, the semiclassical ground-state energy is $-0.565$ (see Eq. (18) below), while the exact numerical result is −0.552 [5]. Thus, the semiclassical energies reproduce well the exact energies even for the ground state. We note that the maximum value of the ground-state energy equals −0.4892 and is attained for $s = 0.8795$. As the exponent approaches $s = 2$, the ground-state energy decreases rapidly to minus infinity, which corresponds to the particle fall to the origin.

## 3. CASE OF $s = 1/2$

For $s = 1/2$, we find from Eq. (17) that $\gamma = -1/6$, and Eq. (16) yields the semiclassical energies of the bound states

$$E_n = -\left(n - \frac{1}{6}\right)^{-2/3}/2. \quad (18)$$

This case is interesting because there exists an exact implicit analytical solution for the energy [5]:

$$\sqrt{2a}H_{a-1}(-\sqrt{2a}) = -H_a(-\sqrt{2a}); \quad a = (-2E_n)^{-3/2}. \quad (19)$$

For $a = 1$, this equation is evidently satisfied identically, but the corresponding eigenfunction equals zero. The eigenfunctions are written in terms of linear combinations of the Hermite function and confluent hypergeometric function [5]. The argument $z = -\sqrt{2a}$ of the Hermite functions $H(z)$ in Eq. (19) is negative and, for $a \gg 1$, falls in the intermediate region, where $z \approx -\sqrt{2\nu + 1}$. An asymptotic expansion suitable for this region is [6]

$$H_\nu(z) = \exp\left(\frac{z^2}{2}\right) 2^{\nu/2+1/4} \pi^{1/4} \nu^{-1/12} \sqrt{\Gamma(\nu+1)}$$
$$\times \{\cos(\pi\nu)\text{Ai}(t) - \sin(\pi\nu)\text{Bi}(t)\}; \quad (20)$$
$$t = -2^{1/2}\nu^{1/6}(z + \sqrt{2\nu+1}),$$

where Ai and Bi are the Airy functions. Using the expansions for the Airy functions in the vicinity of zero (for large values of $a$, the arguments of the Airy functions are close to zero) $t \approx \pm a^{-1/3}/2$ for $\nu = a - 1$ and $\nu = a$, in the limiting case of $a \to \infty$ and taking into account that $\text{Bi}(0) = 3^{1/2}\text{Ai}(0)$, Eq. (19) can be simplified to the form $\cos(\pi a) + \sqrt{3}\sin(\pi a) = 0$, whence we obtain the spectrum given by Eq. (18) with a Maslov index of $-1/6$.

## 4. LIMITING CASES

Let $s = 2 - \beta$, where $\beta \ll 1$. Then, according to general formula (16), we obtain semiclassical energies

$$E_n = -\left(\frac{2\sqrt{2}}{\pi n \beta}\right)^{4/\beta}. \quad (21)$$

It follows from Eq. (21) that the energy levels belong to one of two types: the level energies for $n > \frac{2\sqrt{2}}{\pi\beta}$ tend to zero, while these energies for $n < \frac{2\sqrt{2}}{\pi\beta}$ tend to minus infinity, which indicates the onset of the particle fall to the origin for $s = 2$. There is an abrupt boundary between these two types of levels.

Another limiting case is $s \ll 1$. From Eq. (16), we find

$$E_n = -\left(n - \frac{1}{4}\right)^{-s}. \quad (22)$$

The energies are close to −1, and they begin to rise only for enormous quantum numbers of the order of $n \sim \exp(1/s)$.

Since $-x^{-s} = -\exp(-s \ln x) \approx -1 + s \ln x$ for $s \ll 1$ the Maslov index equals $-1/4$ also for the limiting form of the logarithmic potential $V(x) = V_0 \ln(x/x_0)$ ($V_0 > 0$) according to Eq. (22). The semiclassical energies for this potential obtained by the Bohr–Sommerfeld quantization rule are

$$E_n = V_0 \ln\left[\frac{\hbar}{x_0}\sqrt{\frac{2\pi}{mV_0}}\left(n - \frac{1}{4}\right)\right], \quad (23)$$

where we have written explicitly the Planck constant and the particle mass. These energies can be both positive and negative. Equation (23) can also be derived from Eq. (16) if, for the potential $V(x) = -\lambda x^{-s}$, we pass to the limit $\lambda \to \infty$ and $s \to 0$ with $\lambda s = V_0$. If we take $V_0 = \hbar^2/2mx_0^2 = 1$, Eq. (23) yields $E_1 = \ln(3\sqrt{\pi}/2) = 0.978$ for the semiclassical ground-state energy. The exact numerical value for this case is 1.044 [8], which is in good agreement with the analytical result obtained in the semiclassical approximation.

## 5. SPHERICALLY SYMMETRIC POTENTIAL

The above results can be applied without any modification to states of zero angular momentum in a spherically symmetric potential $V(r) = -V_0/r^s$. A well-known example is the Coulomb potential ($s = 1$), where the Maslov index equals zero and the semiclassical energy coincides with the exact value. Next, let us consider states of a nonzero angular momentum. The radial Schrödinger equation can be written as

$$-\frac{1}{2}\frac{d^2\psi}{dr^2} - \frac{1}{r^s}\psi + \frac{l(l+1)}{2r^2}\psi = E_n\psi; \quad (24)$$
$$n \gg 1, \quad |E_n| \ll 1.$$

For $r \ll r_0 = (-2E_n)^{1/s}$, it becomes the Malmstén equation [7]:

$$-\frac{1}{2}\frac{d^2\psi}{dr^2} - \frac{1}{r^s}\psi + \frac{l(l+1)}{2r^2}\psi = 0, \quad (25)$$

which has the following solution regular at $r = 0$:

$$\psi(r) \sim \sqrt{r} J_\nu\left(\frac{2\sqrt{2}}{2-s} r^{1-s/2}\right); \quad \nu = \frac{2l+1}{2-s}. \quad (26)$$

We note that the singularity in this case lies in the classically forbidden region. The asymptotic behavior of this solution for $r \gg 1$ is

$$\psi(r) \sim r^{s/4} \cos\left(\frac{2^{3/2}}{2-s} r^{1-s/2} - \frac{\nu\pi}{2} - \frac{\pi}{4}\right). \quad (27)$$

Fitting it to the solution given by Eq. (15), which starts from the right turning point, we obtain the following expressions for the eigenvalues and the Maslov index:

$$E_n = -\left\{\frac{\sqrt{2\pi}\Gamma\left(\frac{1}{s}\right)}{\Gamma\left(\frac{1}{s} - \frac{1}{2}\right)}(n+\gamma)\right\}^{\frac{2s}{s-2}}; \quad (28)$$

$$\gamma = \left[\frac{s-1+2l}{2(2-s)}\right]. \quad (29)$$

For $s = 1$, we have, as is expected, $\gamma = 0$. Expressions (28) and (29) were obtained in another way in [8].

## 6. SCATTERING PROBLEM

It is also of interest to consider the problem of semiclassical $s$-wave scattering of a particle with a low (semiclassical) positive energy $E \ll (m/\hbar^2)^{\frac{s}{2-s}} V_0^{\frac{s}{2-s}}$ by the above potential. In much the same way as Eq. (12), let us write the wavefunction in the semiclassical region as

$$\psi(r) \sim \frac{1}{\sqrt{p(r)}} \sin\left(\frac{1}{\hbar}\int_0^r p(r')dr' - \lambda\right);$$

$$p(r) = \sqrt{2m\left(\frac{V_0}{r^s} + \frac{1}{r_0^s}\right)}; \quad r \gg 1. \quad (30)$$

Here, $r_0 = E^{-1/s} \gg 1$ (in dimensionless units). In order to determine $\lambda$, the wavefunctions given by Eqs. (30) and (10) should be matched with each other in the region $1 \ll r \ll r_0$, where both of them are valid:

$$\psi(r) \sim r^{s/4} \sin\left(\frac{2^{3/2}}{2-s} r^{1-s/2} - \frac{s\pi}{4(2-s)}\right). \quad (31)$$

Then, we obtain

$$\lambda = \frac{s\pi}{4(2-s)}. \quad (32)$$

The semiclassical scattering phase is determined as the difference between the phase in Eq. (30) and the phase for the case of vanishing potential (for $1 < s < 2$, because the known logarithmic divergence of the scattering phase exists already for the Coulomb potential ($s = 1$); for $s \leq 1$, all phases for any angular momentum diverge [3]):

$$\delta_0(E,s) = \sqrt{\frac{2mE}{\hbar^2}}\left(\frac{V_0}{E}\right)^{1/s} \int_0^\infty \left[\sqrt{\frac{1}{x^s}+1} - 1\right] dx. \quad (33)$$

Calculating the integral, we find the zeroth scattering phase

$$\delta_0(E,s) = \sqrt{\frac{2mE}{\hbar^2}}\left(\frac{V_0}{E}\right)^{1/s} \frac{\Gamma\left(\frac{2}{s}-1\right)\Gamma\left(1-\frac{1}{s}\right)}{2^{2/s-1}\Gamma\left(\frac{1}{s}\right)} \gg 1, \quad (34)$$

which contributes $\sigma_0 = \pi\hbar^2/mE$ to the cross section. The total cross section is infinite at summation over large orbital angular momenta [3].

The solutions obtained may be useful for the analysis of quark–antiquark bound states [8].

We are grateful to V.V. Kiselev for valuable advice. This study was supported by the Ministry of Education and Science of the Russian Federation (project no. 3.679.2014/K); by the State Committee of Science, Ministry of Education and Science of the Republic of Armenia (grant no. 15T-1C323); and by the program "Leading Research Universities of Russia" (project no. FT_24_2016 for Tomsk Polytechnic University).